# Electronic structure and anomalous band-edge absorption feature in multiferroic MnWO$_4$: An optical spectroscopic study

Woo Seok Choi[1], K. Taniguchi[2], S. J. Moon[1], S. S. A. Seo[1,a)], T. Arima[2], H. Hoang[3], I.-S. Yang[3], T. W. Noh[1], and Y. S. Lee[4,*]

[1]*ReCOE & FPRD, Department of Physics and Astronomy, Seoul National University, Seoul 151-747, Korea*

[2]*Institute of Multidisciplinary Research for Advanced Materials, Tohoku University, Sendai 980-8577, Japan*

[3]*Department of Physics, Ewha Womans University, Seoul, 120-750, Korea*

[4]*Department of Physics, Soongsil University, Seoul 156-743, Korea*

We investigated the electronic structure and lattice dynamics of multiferroic MnWO$_4$ by optical spectroscopy. With variation of polarization, temperature, and magnetic field, we obtained optical responses over a wide range of photon energies. The electronic structure of MnWO$_4$ near to the Fermi level was examined, with inter-band transitions identified in optical conductivity spectra above a band-gap of 2.5 eV. As for the lattice dynamics, we identified all the infrared transverse optical phonon modes available according to the group-theory analysis. Although we did not observe much change in global electronic structure across the phase transition temperatures, an optical absorption at around 2.2 eV showed an evident change depending upon the spin configuration and magnetic field. The behavior of this band-edge absorption indicates that spin-orbit coupling plays an important role in multiferroic MnWO$_4$.

I. INTRODUCTION

Magnetoelectric multiferroic oxides show not only simultaneous orderings of both electric and magnetic order parameters, but also a coupling between them.[1-3] Due to these intriguing physical properties, multiferroic oxides have recently been attracting considerable attentions. In particular, ferroelectricity can be induced through the non-collinear spin structure in several transition metal oxides including $R$MnO$_3$ ($R$ = rare earth ions),[4, 5] Ni$_3$V$_2$O$_8$,[6] CoCr$_2$O$_4$,[7] LiCu$_2$O$_2$,[8] and MnWO$_4$,[9] suggesting a strong magnetoelectric coupling. In these multiferroic oxides, non-collinear spiral spin structure without centro-symmetry may induce macroscopic spontaneous electric polarization. This process works through the inverse of the Dzyaloshinski-Moriya interaction, taking into account the strong spin-orbit coupling (SOC) effect.[5, 10]

MnWO$_4$ is an outstanding multiferroic oxide that exhibits ferroelectricity induced by its spiral spin structure.[9, 11, 12] One of the advantages of studying MnWO$_4$ is that it has only one kind of magnetic ion (Mn$^{2+}$), whereas other multiferroic oxides usually have multiple magnetic ions, thus hindering investigation of their magnetic properties. With the advantage of single magnetic ion, we are provided with a clear window into the electromagnetic coupling behavior and are able to make simple analyses of the magnetic field dependence and SOC effects. Structurally, MnWO$_4$ has a wolframite crystal structure, containing WO$_6$ octahedra.[9, 13] Its magnetic phases are characterized by various antiferromagnetic states such that the first antiferromagnetic phase appears below $T_{N3} \approx 13.5$ K. Successive magnetic phase transitions occur at $T_{N2} \approx 12.5$ K and $T_{N1} \approx 6.5$–8 K, forming three different antiferromagnetically ordered phases: AF1 ($T \leq T_{N1}$), AF2 ($T_{N1} \leq T \leq T_{N2}$), and AF3 ($T_{N2} \leq T \leq T_{N3}$).[14, 15] These magnetic phases have been characterized by neutron scattering experiments, which have shown that AF1 has a collinear up-up-down-down spin structure, AF2 has a non-collinear spiral spin structure, and AF3 has a sinusoidal collinear spin structure.[14, 15] Spontaneous electric polarization occurs at AF2, which suggests that the non-collinear spiral structure might be the origin of the induced ferroelectricity in MnWO$_4$.

Studies on magnetoelectric coupling in MnWO$_4$ have quite recently started to expand their scope to multiple new aspects. The ferroelectric loop and domain dynamics of MnWO$_4$ have been studied under a magnetic field.[16-18] Additionally, suppression of the ferroelectric phase and change in magnetic ordering by doping Fe and Co ions has been observed.[19,20] More recently, the magnetoelectric memory effect was identified in MnWO$_4$ by applying a magnetic field.[21] In this series of works, much attention was paid to the direct observation of the ferroelectric polarization and its behavior under a magnetic field.

Despite such extensive studies on MnWO$_4$, detailed optical spectroscopic investigations have not been reported. Optical spectroscopy is an indispensable tool in studying the multiferroic oxides, for example, the electromagnon may be observed in the THz energy region for $R$MnO$_3$ or $R$Mn$_2$O$_5$, providing direct evidence of electromagnetic coupling.[22, 23] Also, changes in the electronic structure yielded information on the charge-lattice coupling or charge-spin coupling in some oxide multiferroics.[24, 25] Moreover, it should be noted that optical spectroscopy is an excellent tool to investigate various interactions within materials, such as electron charge/spin/lattice/orbital coupling.[26] In particular, given that the SOC is an essential factor in the formation of ferroelectricity by the spiral spin structure in multiferroic oxides, optical investigation is crucial to understand the magnetoelectric coupling in MnWO$_4$.

In this paper, we present an optical spectroscopic study on single crystal MnWO$_4$. By varying light polarization and temperature and by application of a high magnetic field, we characterized the electronic structure and lattice dynamics of MnWO$_4$. The optical spectra of MnWO$_4$ indicate that Mn 3$d$ and W 5$d$ bands contribute to the electronic structure near the Fermi energy, generating a band gap at 2.5 eV. For the lattice dynamics, we recognized all of the 15 IR active phonons calculated from the group-theory analyses. In addition to these features, we identified a tiny optical absorption feature just below the fundamental band gap, which shows a strong temperature and magnetic field dependence. This band-edge absorption feature could be attributed to the intra-atomic $d$-$d$ transition between the Mn states, implying the importance of the SOC in MnWO$_4$.

II. EXPERIMENTS

We synthesized single crystals of MnWO$_4$ using the floating zone method. The crystals were oriented using Laue x-ray photographs and cleaved into thin plates with wide faces perpendicular to the crystallographic principal axis $b$ (010). To obtain the faces perpendicular to $a$ (100) and $c$ (001), the crystals were cut and then polished down to 0.3 $\mu$m using diamond pastes. More details on the single crystal growth have been described elsewhere.[12] The thicknesses of the single crystal were 500, 450, and 700 $\mu$m for (100), (010), and (001) surfaces, respectively.

We measured near-normal incident reflectance ($R(\omega)$) and transmittance ($T(\omega)$) spectra of MnWO$_4$ in a wide photon energy ($\omega$) range (3.7 meV to 20 eV) for each axis. We chose the axes near the crystallographic axes that did not show any oscillation due to birefringence. We used

Fourier-transform infrared spectrophotometers (Bomem DA8 and Bruker IFS66v/S with 4-8 cm$^{-1}$ resolution) between 3.7 meV and 1.5 eV, a near-infrared-visible-ultraviolet grating monochromator (CARY 5G) between 0.6 and 5.9 eV, and an ultraviolet synchrotron spectrometer (3B1 beamline) in the Pohang Light Source (PLS, Korea) from 4.8 eV to 20 eV. The magnetic field (*H*) dependent optical spectra between 1.11 eV (1120 nm) and 3.26 eV (380 nm) were measured using a grating spectrophotometer equipped with a 33 T resistive magnet at the National High Magnetic Field Laboratory (Tallahassee, FL, USA).

III. RESULTS AND DISCUSSION

**A. Electronic structure of MnWO$_4$**

The thin MnWO$_4$ samples had a transparent window between ~0.2 eV and ~2.0 eV, which determines the red color of the single crystal. In this spectral region, we obtained $T(\omega)$ along with $R(\omega)$ and used both to derive complex dielectric constants from a numerical iteration process (the intensity transfer matrix method (ITMM)[27]). To obtain the optical conductivity spectra ($\sigma(\omega)$) over the whole photon energy region, we conducted Kramers-Kronig (KK) analysis using $R(\omega)$. For KK analyses, we extrapolated reflectance spectra in low frequency regions below our measurements as a constant. For frequency regions above 20 eV, we extended reflectance at 20 eV – 30 eV as a constant and then assumed $\omega^{-4}$ dependence.[28, 29] We further confirmed that the results from ITMM and KK analyses using $R(\omega)$ were consistent.[30]

Figure 1 shows $R(\omega)$ for MnWO$_4$ on different axes at 300 K. The electric field of the incident light was polarized to show the optical responses of the crystallographic axes indicated in the figure. We observed that $R(\omega)$ for the (001) and (100) planes, with the light polarized along the *b*-axis, were nearly identical, indicating a high quality of the single crystals and validating our optical experiment. The spectral features below 0.2 eV mostly come from the complex phonon structure in the system, which will be addressed in detail in the next section.

In MnWO$_4$, Mn$^{2+}$ ions have five electrons in the 3*d* orbital with high spin configurations, whereas no electron is present in the 5*d* orbital for W$^{6+}$ ions. For this reason, the unoccupied electronic state of MnWO$_4$ near the Fermi level should be composed of W 5*d* states and Mn 3*d* states, with the occupied state composed of Mn 3*d* states. Due to crystal field splitting of MnO$_6$ and WO$_6$ octahedra, each state is split into $t_{2g}$ and $e_g$ states.

Figure 2 shows $\sigma(\omega)$ of MnWO$_4$ at 300 K obtained from the KK analyses. Besides the low-$\omega$ phonon parts, we could observe larger peaks at higher energies, giving rise to a band gap at

~2.5 eV. We could well describe the higher energy peaks with the sum of Lorentz oscillators as shown as dotted grey lines in Fig. 2. The thin grey lines represent each oscillator. Our result indicates that the electronic structure of $MnWO_4$ consists mainly of two broad peaks located at ~5 and ~6.5 eV. These two peaks might be attributed to a charge transfer excitation from O $2p$ to Mn $3d$ and W $5d$ states, regarding their sizeable strength and very broad nature. There are also small peak like structure at lower energies which could not be fully understood using Lorentz oscillators. These peaks might originate from the inter-band $d$-$d$ transitions, probably from Mn $3d$ to W $5d$ states. Resolution of a more detailed electronic structure and an exact peak assignment would be possible by comparison to the band calculation, which is not currently available for $MnWO_4$.

### B. Optical phonon structure of $MnWO_4$

Focusing on phonon structure, $MnWO_4$ has a wolframite crystal structure that belongs to the monoclinic space group P2/c and $C_{2h}$ point-group, with two formula units per primitive cell. Site group and factor group analyses provide us 18 (8 $A_g$ + 10 $B_g$) Raman active modes and 18 (8 $A_u$ + 10 $B_u$) IR active modes. The detailed analyses are summarized in Table I, while study on the Raman active phonons is summarized elsewhere.[13,31] By excluding the acoustic modes (one $A_u$ and two $B_u$ modes), we had a total of 15 (7 $A_u$ + 8 $B_u$) IR active modes. Figure 3 shows the expanded $\sigma(\omega)$ between 10 and 120 meV at 300 K, where all 15 IR active transverse optical (TO) phonons were identified. The assignments of the experimentally observed TO phonons are summarized in Table II, showing all the phonon modes predicted by the theoretical analyses. Note that the phonon numbers 6, 7, 14 and 15 are observed for more than one axis, suggesting that these phonons are degenerate. Whereas the electronic structures in Fig. 2 were rather isotropic, the phonon structures in Fig. 3 show a striking anisotropy. This might be due to the complex crystal structure of $MnWO_4$ which it determines most of phonon structure but not as much the electronic structure.

It is also noteworthy that the TO phonons only exhibit monotonous changes with decreasing temperature, and no anomalies in their peak positions or spectral weight could be observed across the transition temperatures. In particular, we could not observe a distinct change of the phonon spectra in AF2 phase, possibly due to the limitation of our experimental resolution. For example, temperature dependent behaviors of phonons number 14 and 15 are presented in the inset of Fig. 3(a). The peak position and full width half maximum of the phonon peak show only negligible changes. Such observation might seem to be opposed to the additional reflection appearing below the magnetic phase transition temperature observed by the x-ray diffraction experiment.[12] However, the optical phonon spectra characterize the lattice dynamics and is much less sensitive to the commensurate or incommensurate lattice spacing, as compared to the x-ray diffraction measurement. Moreover, our observation is consistent with the recent Raman studies, where Raman active phonons likewise do not show an anomaly across the magnetic and ferroelectric transition temperatures.[13,31] Therefore, a direct relation between the phonon spectral features and the magnetoelectric coupling properties in $MnWO_4$ could not be identified.

The non-existence of the phonon spectra change across the phase transition temperature is also consistent with the missing electromagnon in $MnWO_4$ in the THz region. The electromagnon is one of the most direct evidences for the multiferroic character having non-collinear spin structure, and it has previously been observed in the FE phase of orthorhombic $RMnO_3$ and $RMn_2O_5$.[22, 23] To search for the electromagnon feature in $MnWO_4$, we measured the reflectance of $MnWO_4$ in the region between 5 and 80 cm$^{-1}$, which corresponds to the region between 0.62 and 9.92 meV, using THz time domain spectroscopy (data not shown). However, we could not observe any structure emerging in the AF2 phase, which corresponds to the FE phase. Rather, we observed several absorption features of unknown origin in this very low-$\omega$ region over the entire temperature range. Such absorption features might have come from lattice defects, or antiferromagnetic resonance remaining up to high temperature, or other collective excitations. A more detailed and systematic optical study on the THz region is required to clearly reveal the low-$\omega$ region spectra.

**C. An anomalous band-edge absorption feature just below the fundamental band gap**

We now turn our attention to a distinct optical absorption observed just below the band gap.[32, 33] As shown in Fig. 4(a), $T(\omega)$ of the $MnWO_4$ (010) plane showed a dip structure at ~2.2 eV, before the fundamental gap at ~ 2.5 eV. This dip resulted in a peak in $\sigma(\omega)$ at ~2.2 eV, as shown in Fig. 4(b). The strength of the peak is very small, with a peak height of less than 1 $\Omega^{-1}$cm$^{-1}$ obtained from the above mentioned ITMM. Such a minute optical feature is only visible in $T(\omega)$, as it is comparatively much more sensitive than is $R(\omega)$. Contrary to the negligible temperature dependence of the overall electronic structure, we could observe a distinct change of the ~2.2 eV peak structure by varying temperature and/or applying a high magnetic field. To analyze the temperature and magnetic field dependence, we calculated optical-absorption spectra $\alpha(\omega)$ from $T(\omega)$ using the simple relation $\alpha(\omega)$ = -log $T(\omega)$ / $d$, where $d$ is the thickness of the sample.[34]

Figure 5(a) shows $\alpha(\omega)$ of the $a$-, $b$-, and $c$-axis responses of $MnWO_4$ at 300 K. For the relative strength of the peak at ~2.2 eV for different polarizations, we note that the $a$-axis response is the strongest, followed by the $c$-axis and then $b$-axis responses. Figures 5(b), 5(c) and 5(d), show the temperature dependence of $\alpha(\omega)$ of the $a$-, $b$-, and $c$-axis responses of MnWO$_4$, respectively. The ~2.2 eV peak becomes sharper with decreasing temperature, down to 8 K, for all polarizations. However, when it enters the AF1 phase ($\leq$ 7K), which is known to be a commensurate collinear spin structure, the peak structure becomes more susceptible to temperature, and its strength decreases quite drastically. Such a distinct change across the magnetic phase transition temperature indicates a strong correlation between the magnetic spin structure and electronic structure in MnWO$_4$.

In addition to the temperature dependence, we noticed unusual magnetic field dependence for the ~2.2 eV peak. A high magnetic field of 30 T was applied along $H // c$ at 4, 9, and 15 K, which correspond to the AF1, AF2, and PM phases of MnWO$_4$, respectively. Figure 6 shows the magnetic field-dependent $\alpha(\omega)$ for MnWO$_4$. Here, on the (010) plane, MnWO$_4$ was measured without a polarizer, and we obtained a mixed optical response of the $a$-axis and $c$-axis. The spectral feature of the ~2.2 eV peak is invariant with increasing magnetic field strength up to 30 T, in both the PM and AF2 phases. However, in the AF1 phase, we observed a distinct change. The decreased peak intensity in the AF1 phase at 0 T is recovered to that observed at higher temperatures (or other magnetic phases) with increasing magnetic field strength. The critical magnetic field strength required to recover the spectral features for the AF2 or PM phases was around 17 T (Ref. 35).

One possibility for the origin of the ~2.2 eV peak observed for MnWO$_4$ is the exciton formed in a polar crystal. An exciton is a hydrogen-like bound state of an electron and hole arising from a long-range Coulomb interaction. It usually appears just below the band gap in optical spectra and is due to the impurity levels located between the conduction band and the valence band. In addition, the typical strength of the quasi-particle is comparable to that observed for the ~2.2 eV peak.[36] However, the weak temperature and polarization dependence cannot be explained with normal excitons. In addition, the peak profile is quite symmetric, contrary to normal exciton peaks, which have an asymmetric peak profile.

Instead of the exciton model, it seems more plausible to attribute the ~2.2 eV peak to the intra-atomic $d$-$d$ transition assisted by the SOC. Because the optical transition is basically an electric dipole transition, the $d$-$d$ transition is originally orbital forbidden, considering its parity.[37, 38] In addition, the intra-atomic $d$-$d$ transition in the Mn site is also spin forbidden, meaning that only transitions between states with the same spin are allowed. Given that Mn$^{2+}$ in MnWO$_4$ has a high spin configuration with five $d$ electrons in the same spin orientation, the transition to the opposite unoccupied spin state is normally impossible. However, the orbital selection rule could be partially broken due to the non-inversion centric position of the Mn ion within the crystal structure of MnWO$_4$. In this case, parity is no longer a good quantum number, and the even and odd states can become partially mixed.[38] Moreover, the spin forbidden nature of the *d-d* transition could be overcome by the SOC. Through SOC, the spin directions become mixed with each other, modifying the spin states.[39] Although SOC in 3*d* orbitals is generally not very strong, it can play an important role in multiferroic oxides such as *R*MnO$_3$ or MnWO$_4$. Indeed, in these multiferroic manganites with non-collinear spin structure, ferroelectricity may be understood through a microscopic model involving SOC.[10] Due to this SOC, the high spin configuration is disturbed such that the intra-atomic *d-d* transition becomes possible with a disordered/non-collinear spin configuration. A similar observation of the ~2.2 eV peak has been made by Nogami *et al.* and also by Ejima *et al.*, where both of them also attributed the peak to the intra-atomic *d-d* transition within the Mn$^{2+}$ ion.[32, 33]

Attribution of the ~2.2 eV peak to the intra-atomic *d-d* transition provides a clue in explaining our temperature- and magnetic field-dependent results. In particular, focusing on the drastically decreased peak intensity at the AF2-AF1 transition temperature, we suggest that the formation of a long-range commensurate collinear spin structure in the AF1 phase would suppress the SOC-allowed transition. Because the spin structure for the AF1 phase is an up-up-down-down type, the SOC-allowed optical transition peak intensity would be decreased due to the rigid spin directions. However, when a high magnetic field is applied or the temperature is raised, the collinear antiferromagnetic spin structure would be destroyed, so that SOC may play a role in restoring the ~2.2 eV peak intensity to that observed for the higher temperature phases. In addition, the increasing intensity of the ~2.2 eV peak in the order of *a*-, *c*-, to *b*-axis suggests that the SOC negates the spin-forbidden nature of intra-atomic *d-d* transition most strongly for the configuration where the electric field of the light is applied in the spin direction.

The ~2.2 eV peak did not show a significant change across the AF2-AF3 and AF3-PM phase transitions. This suggests that macroscopic electric polarization is irrelevant to the ~2.2 eV peak structure. The distinction of the AF1-AF2 phase transition may be related to the observation that only this transition is of first order, whereas the other transitions are of second order.[14] The associated energy of the AF1-AF2 phase transition may have an influence on the ~2.2 eV peak. On the other hand, our results may signify that the local AF3 phase is not actually collinear, in that the band-edge absorption peak intensity in AF3 is stronger than that in AF1. Although the averaged magnetic structure is collinear, as observed in the neutron diffraction experiments,[14] the actual coupling of the spins to the photons may occur on a much shorter time scale.

## IV. SUMMARY

We identified the electronic and IR phonon structures of single-crystal multiferroic $MnWO_4$ by optical spectroscopy. Two main peaks were observed above the band gap that could be attributed to charge transfer transitions between the oxygen and transition metal states. We identified 15 IR active phonon modes predicted by the group theoretical analyses. Although the total electronic structure was quite isotropic, the phonon structure showed a large anisotropy. Just below the fundamental band gap, we additionally observed a band-edge optical transition peak. From the polarization, temperature, and magnetic field-dependent evolution of this peak, we could confirm that the peak represented an intra-atomic *d-d* transition between the Mn states in $MnWO_4$. Our results suggest that the multiferroic oxide $MnWO_4$ involves strong spin-orbit coupling that is responsible for the magnetoelectric coupling.

ACKNOWLEDGMENTS

The authors are grateful to J.-Y. Kim, R. P. Prasankumar, K. S. Burch, and J. W. Kim for valuable discussions. This research was supported by the Basic Science Research Program through the National Research Foundation of Korea (NRF), funded by the Ministry of Education, Science and Technology (MEST) (No. 2009-0080567). The experiments at PLS were supported in part by MEST and POSTECH. A portion of this work was performed at the National High Magnetic Field Laboratory, which is supported by NSF Cooperative Agreement No. DMR-0084173, by the State of Florida, and by the DOE.

FIG. 1. (Color online) $R'(\omega)$ of MnWO$_4$ at 300 K for (a) *a*-axis, (b) *b*-axis, and (c) *c*-axis responses. For a detailed procedure for obtaining $R'(\omega)$, see the text.

FIG. 2. (Color online) $\sigma(\omega)$ of MnWO$_4$ at 300 K for (a) *a*-axis, (b) *b*-axis, and (c) *c*-axis responses in the high-$\omega$ region. The dotted grey lines indicate the result of Lorentz oscillator fitting. The solid grey lines represent each oscillator. The triangles indicate the peak positions of the Lorentz oscillators.

FIG. 3. (Color online) $\sigma(\omega)$ of MnWO$_4$ at 300 K for (a) *a*-axis, (b) *b*-axis, and (c) *c*-axis responses in the low-$\omega$ region. The inset in (a) shows the temperature dependent positions and full width half maximums (FWHM) for the phonon peak for phonons of number 14 (circle) and 15 (square). The size of the symbols represents the error bar.

FIG. 4. (Color online) (a) $T(\omega)$ and (b) $\sigma(\omega)$ of MnWO$_4$ at 300 K for (010) plane in the transparent region.

FIG. 5. (Color online) (a) Optical-absorption spectra of the ~2.2 eV peak structure at 300 K for each axis. Temperature-dependence of the structure is also shown for (b) *a*-axis, (c) *b*-axis, and (d) *c*-axis responses.

FIG. 6. (Color online) Magnetic field-dependent optical-absorption spectra of the ~2.2 eV peak structure of MnWO$_4$ (010) plane. The spectra are measured at (a) 4 K, (b) 9 K, and (c) 15 K, which correspond to the AF2, AF1, and PM phases, respectively.

TABLE I. Factor group analyses for MnWO$_4$ with P2/c space group and C$_{2h}$ point-group.

| Atom | Position | Site symmetry | Raman active | IR active |
|---|---|---|---|---|
| 2W | 2e | C$_2$(2) | 1A$_g$+2B$_g$ | 1A$_u$+2B$_u$ |
| 2Mn | 2f | C$_2$(2) | 1A$_g$+2B$_g$ | 1A$_u$+2B$_u$ |
| 8O1 | 4g | C$_1$(4) | 6A$_g$+6B$_g$ | 6A$_u$+6B$_u$ |
| MnWO$_4$ | 12 | | 8A$_g$+10B$_g$ | 8A$_u$+10B$_u$ |

TABLE II. Phonon peaks in meV for each crystallographic axis.

| No. | *a*-axis | *b*-axis | *c*-axis |
|---|---|---|---|
| 1 | 17 | | |
| 2 | | | 19 |
| 3 | | 21 | |
| 4 | 25 | | |
| 5 | | | 26 |
| 6 | 30 | | 29 |
| 7 | | 38 | 38 |
| 8 | | 42 | |
| 9 | 45 | | |
| 10 | | 52 | |
| 11 | 56 | | |
| 12 | | 62 | |
| 13 | 65 | | |
| 14 | 83 | 82 | 81 |
| 15 | 105 | 107 | 106 |

a) Present address: Materials Sciences and Technology Division, Oak Ridge National Laboratory, Oak Ridge, TN 37831, USA

* Electronic address: ylee@ssu.ac.kr

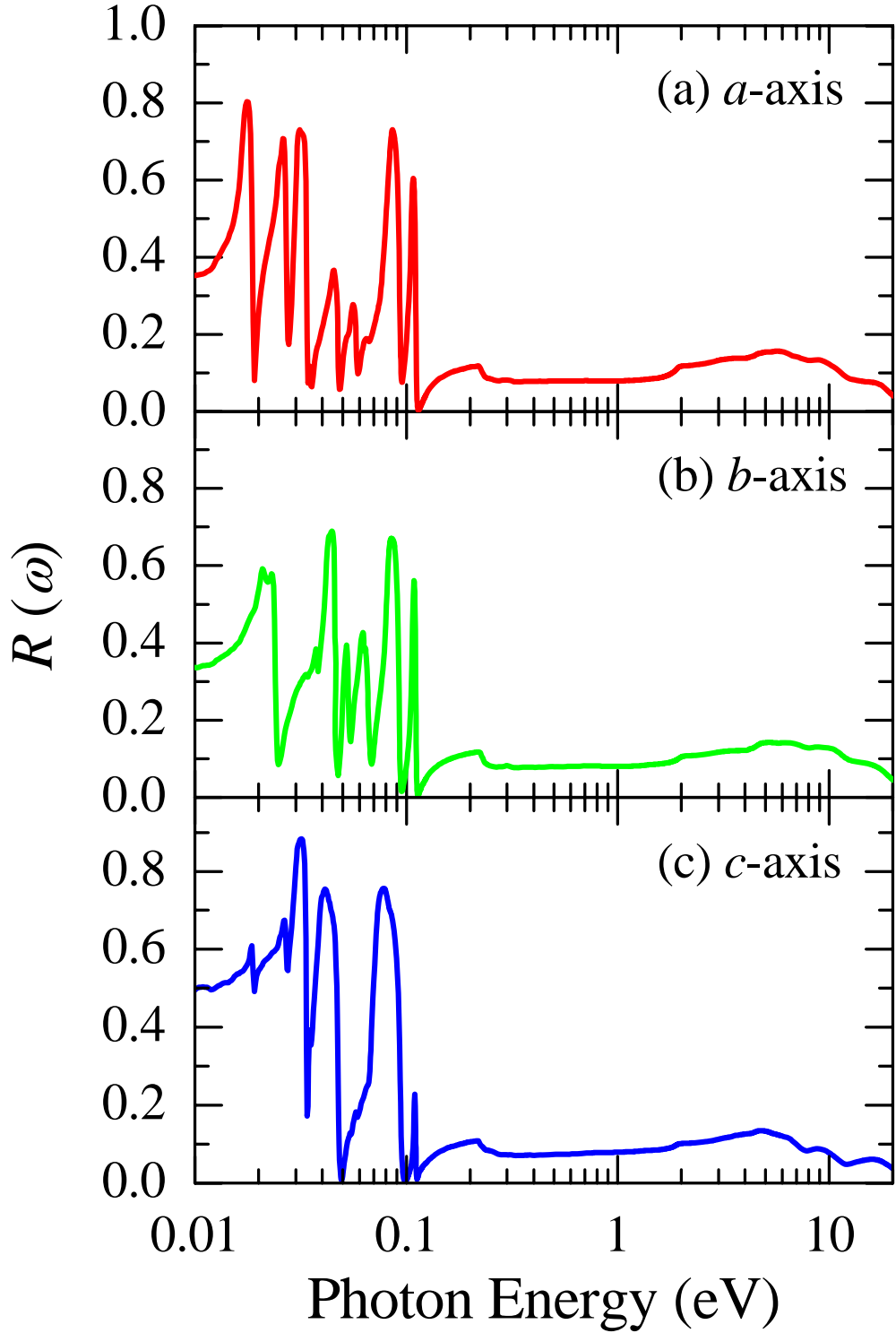

Figure 1
Choi *et al.*

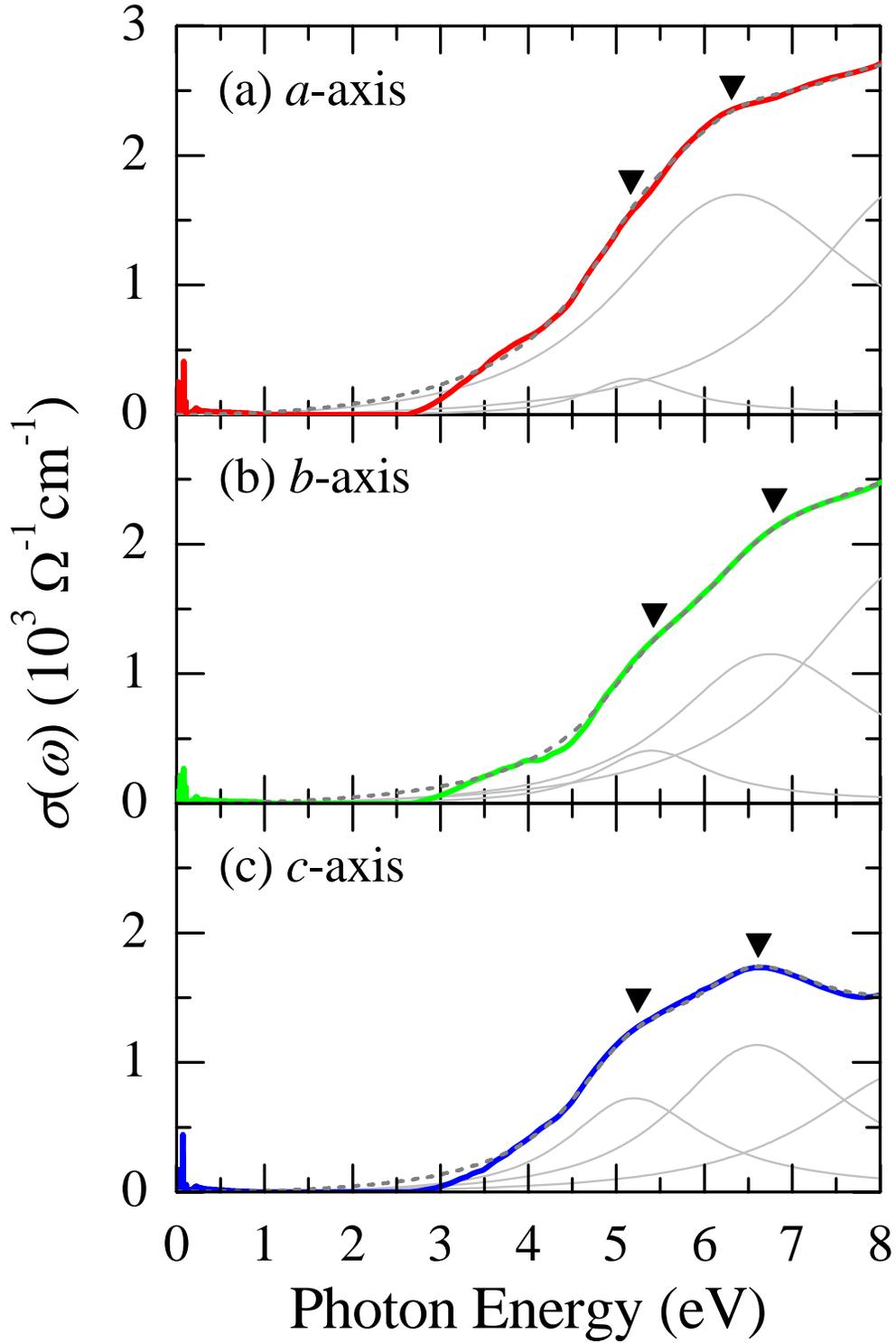

Figure 2
Choi *et al*.

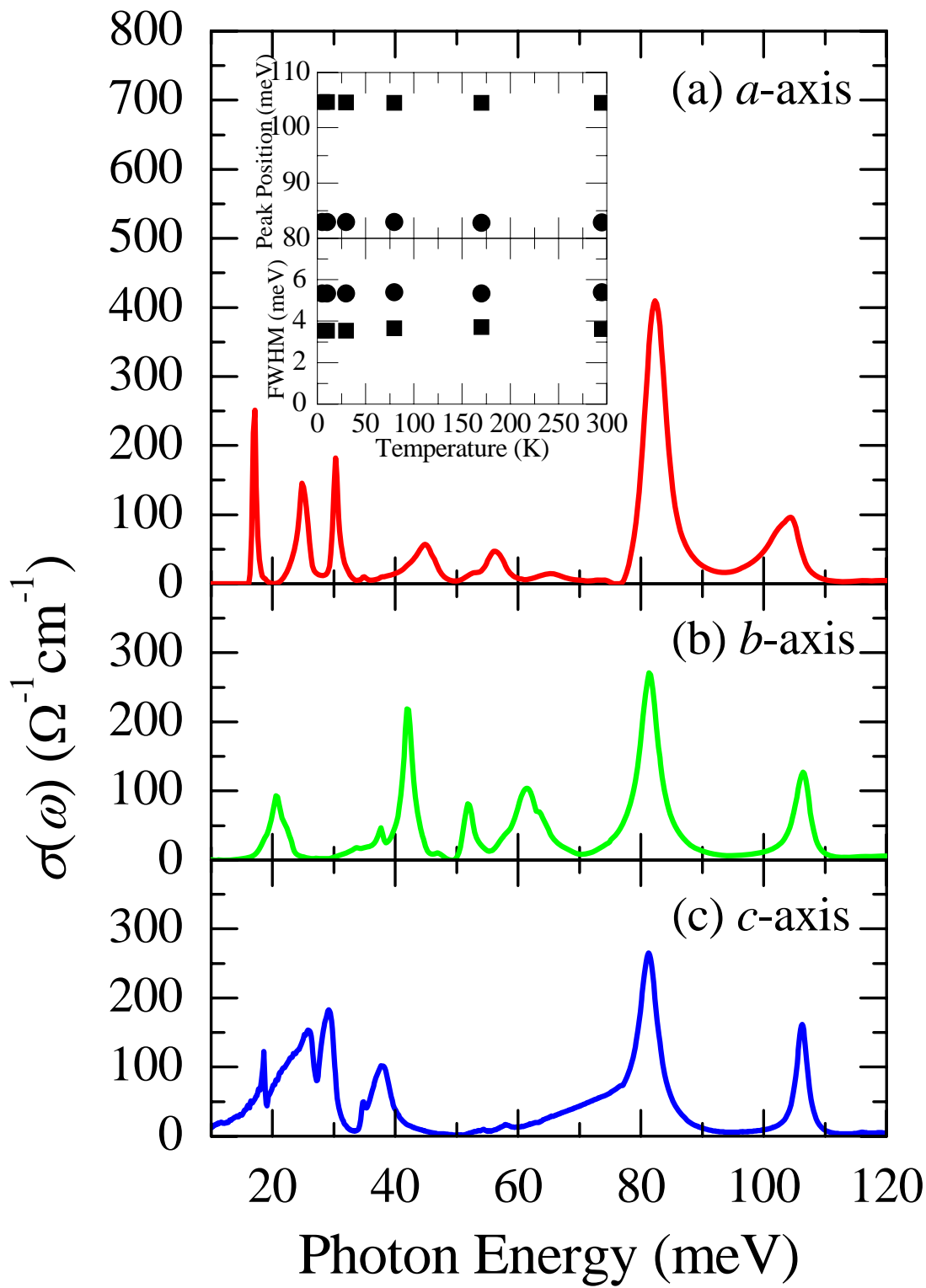

Figure 3
Choi *et al.*

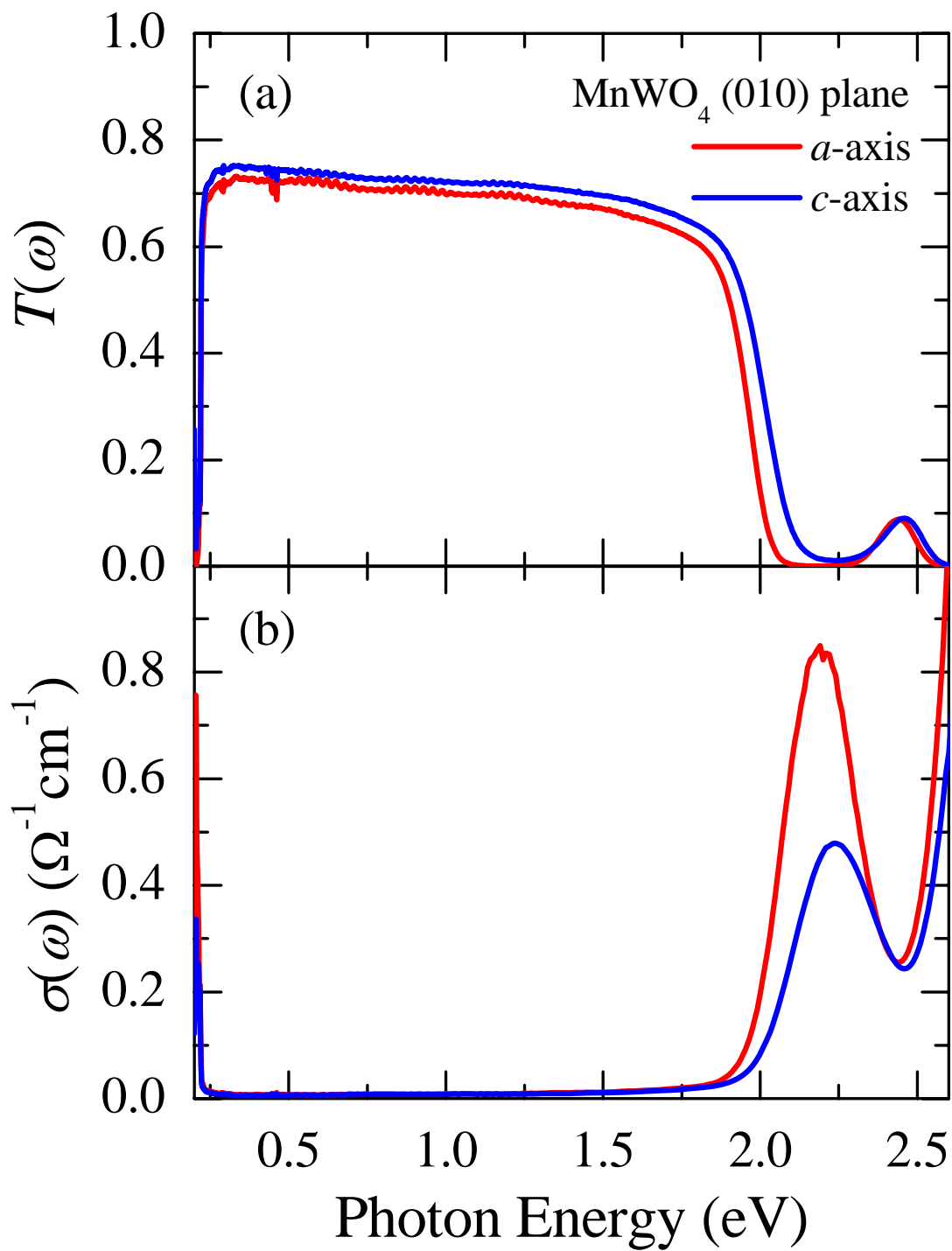

Figure 4
Choi *et al.*

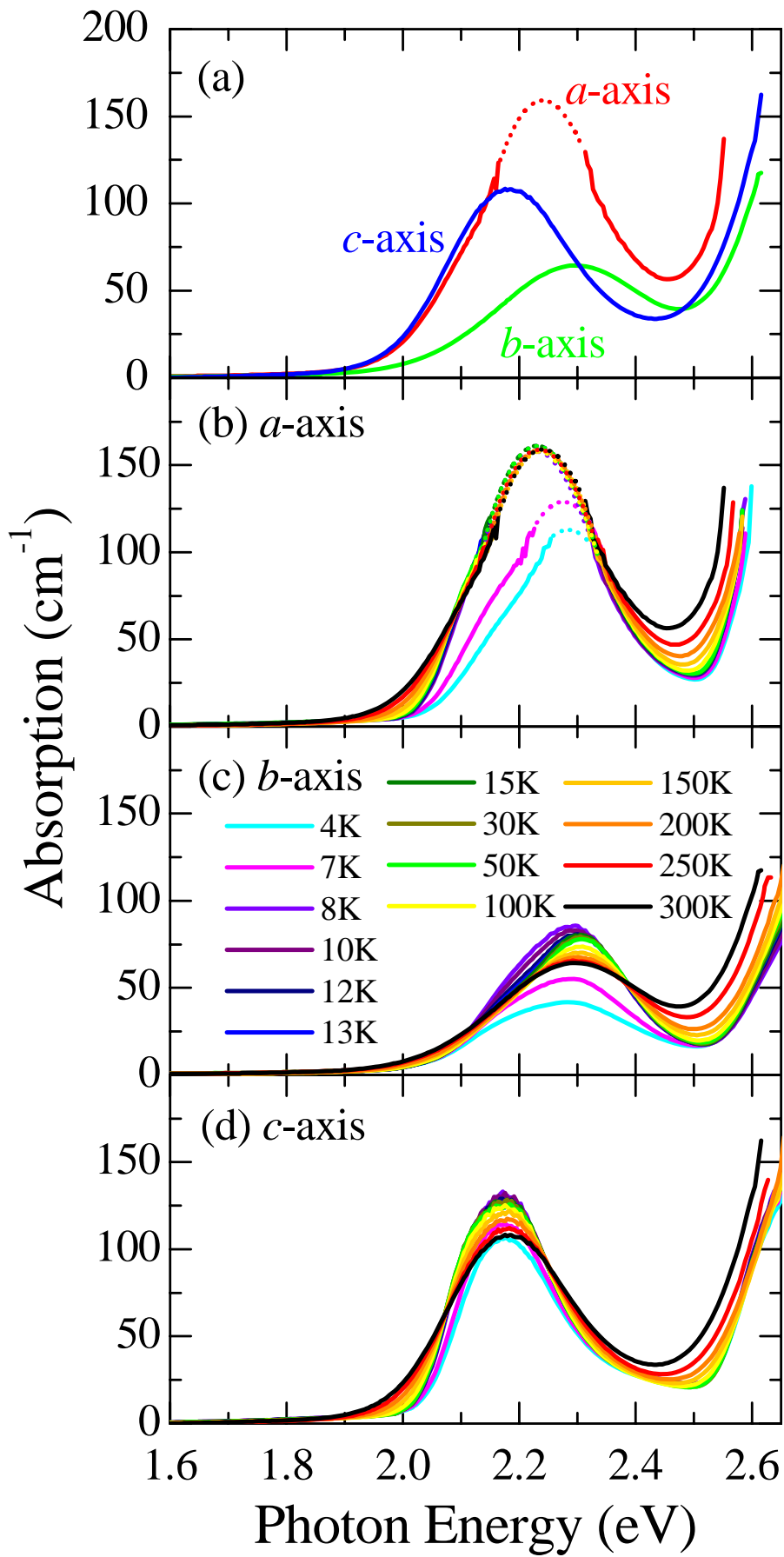

Figure 5
Choi *et al.*

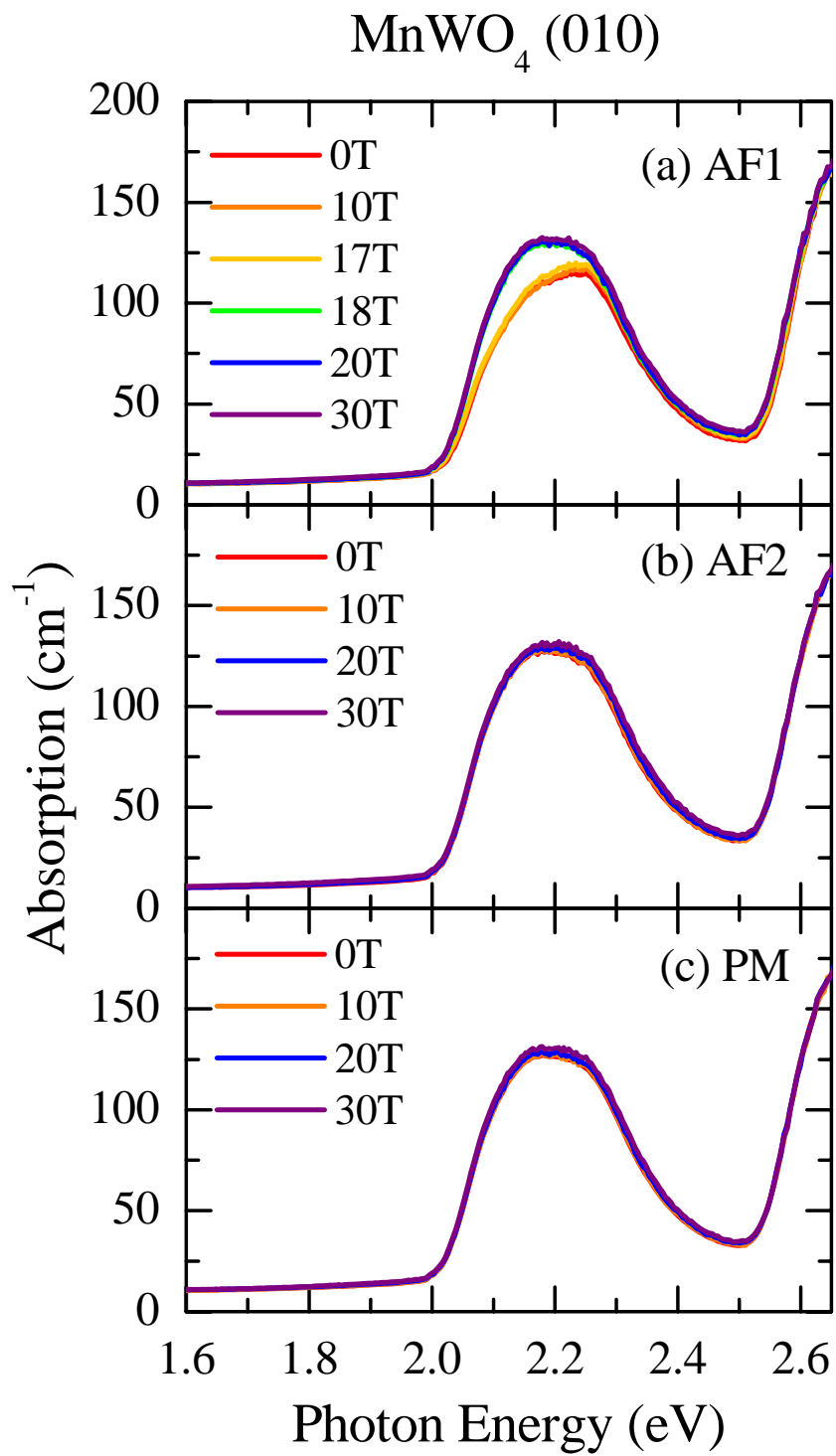

Figure 6
Choi *et al.*